\documentclass{article}
\usepackage{hyperref,amssymb,amsmath,mathrsfs,bm,graphicx}
\usepackage{color}
\usepackage{epsfig,graphics}
\begin{document}
\centerline{\large\bf Self--similar and charged spheres}
\centerline{\large\bf in the free--streaming approximation}
\vspace*{0.245truein}
\centerline
{\footnotesize{\large W. Barreto\footnote{Centro de F\'\i sica Fundamental,
Facultad de Ciencias, Universidad de Los Andes, M\'erida, Venezuela}
and L. Rosales\footnote{Laboratorio de F\'\i sica Computacional, Universidad Experimental
Polit\'ecnica ``Antonio Jos\'e de Sucre'', Puerto Ordaz, Venezuela}}}
\baselineskip=12pt
\vspace*{0.21truein}
\date{\today}
\begin{abstract}
We evolve nonadiabatic charged spherical distributions of matter. 
Dissipation is described by the free--streaming approximation.
We match a self--similar interior solution with the Reissner--Nordstr\"om--Vaidya exterior solution. The transport mechanism is decisive to the fate of the gravitational collapse.
Almost a half of the total initial mass is radiated away. The transport mechanism determines the way in which the electric charge is redistributed.

\vspace*{0.21truein}
\end{abstract}

\renewcommand{\baselinestretch}{2}
\section{Introduction}
There is a renewed interest in the study of self--gravitating spherically
symmetric charged fluid distributions (\cite{dhlms07}, \cite{rbpr10} and references therein).
In self-gravitating systems the electric charge is believed
to be constrained by the fact that the resulting electric field
should not exceed the critical field for pair creation,
$10^{16}$Vcm$^{-1}$ \cite{b71}. This restriction in the critical field has
been questioned \cite{rebels1}--\cite{rebels4} and does not apply to phases of
intense dynamical activity with time scales of the order of
(or even smaller than) the hydrostatic time scale, and for
which the quasistatic approximation \cite{hd97} is clearly not reliable as in the collapse
of very massive stars or the quick collapse phase preceding
neutron star formation.
Electric charge has been studied mostly under static
conditions \cite{i02}--\cite{remlz03}.
Of recent interest are charged quasi--black
holes \cite{lemos} and the electrically charged extension to
quasispherical realization \cite{bonnor}. Distributions electrically
charged can be considered in practice as anisotropic
\cite{brrs07}, \cite{ivanov}. Some authors combine anisotropy and electric charge
 \cite{dhlms07}, \cite{maharaj}, \cite{ma10} but not as a single entity by means of an equation of state.

The electric field has been postulated to be very high in
strange stars with quark matter \cite{u04}, \cite{mh04}, although other
authors suggest that strange stars would not need a large
electrical field \cite{jrs06}. The effects of dissipation, in both
limiting cases of radiative transport, within the context of
the quasistatic approximation, have been studied in \cite{hs03}. Using this approximation
is very sensible because the hydrostatic time scale
is very small, compared with stellar lifetimes, for many
phases of the life of a star. It is of the order of 27 minutes
for the sun, 4.5 seconds for a white dwarf, and 104 seconds
for a neutron star of one solar mass and 10 km radius
\cite{books1}, \cite{books2}. However, such an approximation does not apply
to the very dynamic phases mentioned before. In those
cases it is mandatory to take into account terms which
describe departure from equilibrium, i.e., a full dynamic
description has to be used \cite{hs04}.

We are concerned in this paper with configurations out
of (static) equilibrium with intense dynamical activity. 
We use the approach of considering the electrically charged matter distribution 
as an anisotropic fluid. It is well known that different energy--momentum tensors
can lead to the same spacetime \cite{tt73}--\cite{rs83}. Under spherical symmetry, electric charge can be considered as a special case of anisotropy \cite{brrs07}, \cite{rbpr10}. 
To obtain a dynamical model we consider the free--streaming
approximation as the transport mechanism, and a self--similar spacetime for the
inner region. We explore the fate of the gravitational collapse.
Since dissipation is introduced by the free--streaming of radiation and local anisotropy by the electric charge, it is not necessary to consider in our context any hyperbolic theory of dissipation as for heat flow or viscosity. In many circumstances the mean free path of particles transporting energy may be large enough to justify the free--streaming approximation. We do a comparison with previous work
\cite{brrs07} in which the transport mechanism was diffusive.

Is it well known that the Einstein field equations admit homothetic motion \cite{ct71}--\cite{lz90}. Applications of self--similarity range from modeling black holes to producing
counterexamples to the cosmic censorship conjecture \cite{ch74}--\cite{cc05}. It is well
established that in the critical gravitational collapse of an scalar field
the spacetime can be self--similar \cite{c93}--\cite{g99}. 
We have applied characteristic methods to study the self--similar collapse 
of spherical matter and charged distributions \cite{bd96}--\cite{bd99},\cite{brrs07}.
 The assumption of self--similarity reduces 
the problem to a system of ODE's, subject to boundary conditions determined by matching to 
an exterior Reissner--Nordstr\"om--Vaidya solution. Heat flow in the internal fluid is balanced at 
the surface by the Vaidya radiation. One simulation \cite{bd99} illustrates how a nonzero total charge can 
halt gravitational collapse and produce a final stable equilibrium. It is interesting that the 
pressure vanishes in the final equilibrium state so that hydrostatic support is completely supplied 
by Coulomb repulsion. Another possible final state is extremely compact and 
oscillatory with non zero pressure \cite{brrs07}. In this last case
electric charge redistribution in the fluid is possible.
We explore here if these results depend on the mechanism of transport.

This work is organized as follows. In Sec. 2 we write the field equations for electrically charged interior fluids as seen by Bondian observers. We also show in this section the junction conditions with the exterior spacetime and sketch the general procedure to get physical variables. In Sec. 3 the set of surface equations are presented for parametrized self--similar solutions. Modeling is performed in section 4 to discuss results and conclude in section 5.

\section{Field equations}
We proceed now to describe the matter distribution, the inner and outer line elements and the
field equations.
\subsection{Interior spacetime}
Our starting point is the Bondi approach to study the evolution of gravitating spheres \cite{b64}.
Let us consider a nonstatic distribution of matter which is spherically symmetric. In 
radiation coordinates \cite{bvm62} the metric takes the form
\begin{equation}
ds^2=e^{2\beta}[(V/r)du^2+2dudr] -r^2(d\theta ^2+\sin^2\theta\,d\phi ^2), 
\label{interior}
\end{equation} 
where $\beta$ and $V$ are functions of $u$ and $r$. Here, $u\equiv x^0$ is the
Bondi time at ${\mathcal I}^+$, with $u=$const. on the outgoing cones. $r\equiv x^1$ is
the surface--area distance on these null cones. $\theta\equiv x^2$ and $\phi\equiv x^3$
is the usual polar coordinates. We use geometrized units ($c=G=1$). To get physical
input we use Bondian observers \cite{b09}, introducing the Minkowski
coordinates $(T,X,Y,Z)$ by
\begin{eqnarray}
dT&=&e^\beta[(V/r)^{1/2}du + (r/V)^{1/2}dr],\\
dX&=&e^\beta(r/V)^{1/2}dr,\\
dY&=&r d\theta,\\
dZ&=&r\sin\theta d\phi.
\end{eqnarray}
Next one assumes that for an observer moving relative to these coordinates with
velocity $\omega$ in the radial direction, the space contains a charged fluid
with energy density $\hat\rho$, pressure $\hat p$, electric energy density $\hat\mu$
and unpolarized energy density $\hat\epsilon$ traveling in the radial direction.
For this comoving
observer, the covariant energy tensor is 
\begin{eqnarray}
\left(
  \begin{array}{cccc}
     \hat\rho+\hat\mu+\hat\epsilon & -\hat\epsilon  & 0 & 0\\
     -\hat\epsilon & \hat p -\hat\mu +\hat\epsilon & 0 & 0 \\
      0 & 0 & \hat p + \hat\mu & 0 \\
      0 & 0 & 0 & \hat p + \hat\mu
  \end{array}
\right), 
\end{eqnarray}
where $\hat\mu=E^2/8\pi$ and $E=Q/r^2$ is the local electric field being $Q(u,r)$
interpreted naturally as the charge within the radius $r$ at time $u$, which satisfies
the conservation equation
\begin{equation}
DQ\equiv Q_{,u} + \frac{dr}{du} Q_{,r} =0,
\label{ce}
\end{equation}
where the a comma represents partial derivative respect to the indicated coordinate, and
the matter velocity is given by
\begin{equation}
\frac{dr}{du}=\frac{V}{r}\frac{\omega}{1-\omega}.
\label{mv}
\end{equation}
Observe that the adiabatic fluid has a diagonal covariant energy  tensor $(\rho,p_r,p_t,p_t)$
with $\rho=\hat\rho+\hat\mu$, $p_r=\hat p-\hat\mu$ and
$p_t=\hat p+\hat\mu$, which is exactly the same as for an anisotropic fluid \cite{cosenzaetal}. Clearly the 
electric charge produces local anisotropy, contributing to the matter energy density
and pressure. If electric charge is zero we recover the Pascalian character of neutral matter, that is,
isotropy.

Now, we define the mass function as
\begin{equation}
\tilde m=\frac{1}{2}(r-Ve^{-2\beta}),
\label{mass}
\end{equation}
related with the usual total mass by means of \cite{b71},\cite{mashhoon}
\begin{equation}
m_T=\tilde m + \frac{Q^2}{2r},
\end{equation}
which is the generalization of the Misner--Sharp mass for the charged case \cite{dhlms07}.
Thus, the field equations can be written as \cite{brrs07},\cite{b93}
\begin{equation}
\frac{\rho + p_r\omega^2}{1-\omega^2}+\hat\epsilon\frac{(1+\omega)}{(1-\omega)}=
-\frac{e^{-2\beta}\tilde m_{,u}}{4\pi r(r-2\tilde m)} +\frac{\tilde m_{,r}}{4\pi r^2},
\label{one}
\end{equation}

\begin{equation}
\tilde \rho=\frac{\tilde m_{,r}}{4\pi r^2},
\label{two}
\end{equation}

\begin{equation}
\tilde\rho + \tilde p=\frac{\beta_{,r}}{2\pi r^2}(r-2\tilde m),
\label{three}
\end{equation} 
 
\begin{eqnarray}
p_t=&-&\frac{1}{4\pi}\beta_{,ur}e^{-2\beta}+\frac{1}{8\pi}(1-2\tilde m/r)(2\beta_{,rr}+4\beta_{,r}^2-\beta_{,r}/r)\nonumber\\
&+&\frac{1}{8\pi r}[3\beta_{,r}(1-2\tilde m_{,r})-\tilde m_{,rr}],
\label{four}
\end{eqnarray}
where
\begin{equation}
\tilde p=\frac{p_r-\omega\rho}{1+\omega},
\end{equation}
and
\begin{equation}
\tilde \rho=\frac{\rho-\omega p_r}{1+\omega}.
\end{equation}
Observe that the field equations (\ref{one})--(\ref{four}) are exactly the same as for
anisotropic matter \cite{b93},\cite{hs}, that is, electric charge can be interpreted as local
anisotropy of the fluid. That is possible because of the mass definition given by (\ref{mass}),
otherwise electric charge has to be viewed as part of the metric. From this point of
view, electric charge is formally an additional physical variable which contributes
clearly to the matter energy density and pressure. Also this interpretation
is better understood physically due to the Bondi point of view about the comoving reference system.
This procedure, except for the mass definition, was the same followed to interpret
viscosity as anisotropy \cite{b93}.
\vskip 0.5cm

\subsection{The exterior spacetime and junction conditions}
We consider that the spherically symmetric distribution of collapsing charged fluid
is bounded by the surface $\Sigma$. Outside $\Sigma$ we have the 
Reissner--Nordstr\"om--Vaidya spacetime, that is, all outgoing radiation is massless,
described by
\begin{eqnarray}
ds^2&=&(1-2M(u)/r+q^2/r^2)du^2+2dudr \nonumber\\
&-&r^2(d\theta ^2+\sin^2\theta\,d\phi ^2), 
\label{exterior}
\end{eqnarray}
where $M(u)$ and $q$ denote the total mass and charge, respectively.

It can be shown that the junction conditions for the smooth matching of (\ref{interior}) and
(\ref{exterior}) on $\Sigma$ implies \cite{medinaetal}
\begin{equation}
\hat p \overset{\Sigma}{=} 0,\,\,\,Q \overset{\Sigma}{=} q,\,\,\, m(u,r)\overset{\Sigma}{=} M(u),\,\,\,
\beta\overset{\Sigma}{=} 0,
\label{bc}
\end{equation}
where $\overset{\Sigma}{=}$ means that both sides of equation are evaluated on $\Sigma$. It is remarkable
that the zero pressure on $\Sigma$ is equivalent to the continuity of the second
differential form equation
\begin{equation}
-\beta _{,u}e^{-2\beta}+\left(1-\frac{2\tilde m}{r}\right)\beta_{,r}-\frac{\tilde m_{,r}}{2r}+
\frac{Q^{2}}{4r^{3}}\overset{\Sigma}{=} 0.
\label{sff}
\end{equation}

Up to this point, within the Bondi framework and spherical symmetry, all the written equations
are general. We have five physical variables ($\rho$, $p$, $\omega$, $\epsilon$, $Q$) and
two metric functions ($\tilde m$, $\beta$), for which we have the five
equations (\ref{ce}), (\ref{one})--(\ref{four}). Thus, additional assumptions are necessary to solve the 
characteristic initial value problem. In the next section  we suppose that the spacetime
is self--similar to illustrate how the approach works and explore the influence of the transport mechanism.
But before any assumption to get the metric functions we sketch the general procedure to obtain physical variables.
\subsection{Getting physical variables}
The algorithm to calculate the physical variables once we get the dynamics at $\Sigma$ and the metric functions everywhere
is as follows:
\begin{enumerate}
\item Specifying an initial charge distribution (any) we calculate all physical variables ($\hat p$, $\hat\rho$, 
$\omega$, $\hat\epsilon$) at any piece of material;

\item We integrate numerically equation (\ref{ce}) to advance $Q$ in time;

\item Once again we get all physical variables up to the integrate time at any piece of material.
\end{enumerate}

\section{Self--similarity and surface equations}
{
Self--similarity is invariably defined by the existence of a homothetic
Killing vector field \cite{ct71}. 
A homothetic vector field on the manifold is one that satisfies 
$\pounds_\xi {\bf g}=$2$n{\bf g}$ on a local chart,
 where $n$ is a constant on the 
manifold and $\pounds$ denotes the Lie derivative operator. If $n \ne 0$ we
have a proper homothetic vector field and it can always be scaled to have
$n = 1$; if $n = 0$ then $\xi$ is a Killing vector on the manifold 
\cite{h88,h90,c94}. So, for a constant rescaling, $\xi$ satisfies $
\pounds_\xi{\bf g}=$2${\bf g}$ and has the form 
$
\xi =\Lambda (u,r)\partial_u  +\lambda (u,r)\partial_r
$. 
If the matter field is a perfect fluid, the only
equation of state consistent with $\pounds_\xi{\bf g}=$2${\bf g}$
is a barotropic one \cite{ct71}. The homo\-thetic
 equations reduce to $\xi(X)=0$, $\xi(Y)=0$, $\lambda=r$ and
$\Lambda=\Lambda(u)$, where $X\equiv\tilde m/r$ and $Y\equiv\Lambda e^{2\beta}/r$.
Therefore, $X=X(\zeta)$ and $Y=Y(\zeta)$ are solutions if the self--similar
 variable is defined as $\zeta\equiv r\, \exp(- \int du/\Lambda)$. 
 Here we assume that
 $X=C_1\zeta^k$ and that $Y=C_2\zeta^l$, where $C_1$, $C_2$, $k$ and $l$ are
constants. This power--law dependence on $\zeta$ is based on
the fact that any  function of
$\zeta$ is solution of $\pounds_\xi {\bf g}=$2${\bf g}$.
Demanding continuity of the first fundamental form we get
the following metric solutions \cite{bpr99}, \cite{brrs07}:}
\begin{equation}
\tilde m=\tilde m_\Sigma \left(\frac{r}{r_\Sigma}\right)^{k+1}
\label{eme}
\end{equation}
and
\begin{equation}
e^{2\beta}=\left(\frac{r}{r_\Sigma}\right)^{l+1},
\label{beta}
\end{equation}
where $k$ and $l$ are constants; the subscript indicates that the quantity is evaluated at the surface $\Sigma$.
Thus $r_\Sigma(u)$ represents the radius of the distribution.
\begin{figure}[htbp!]
\begin{center}
\scalebox{.6}{\includegraphics[angle=0]{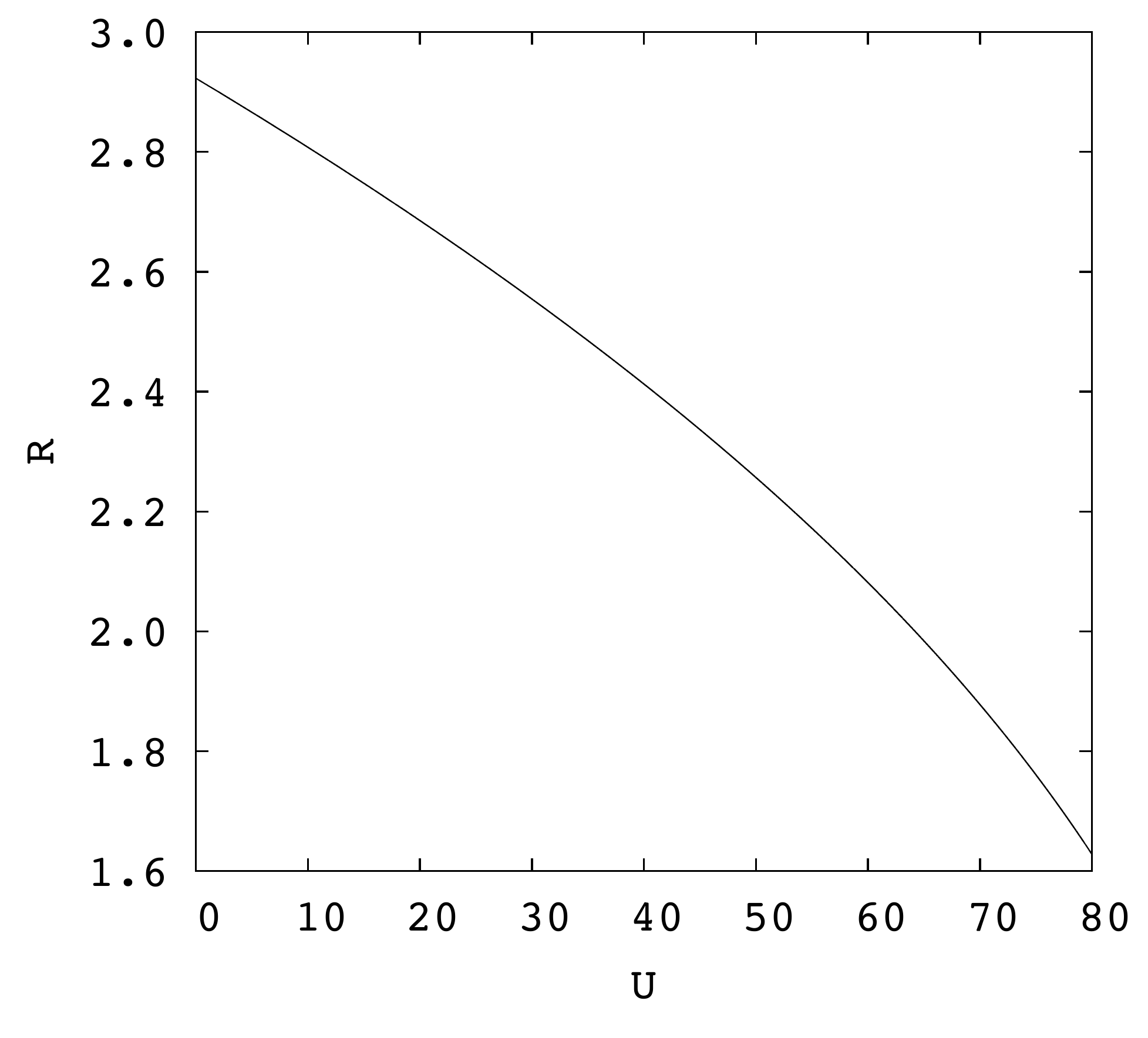}}
\caption{Evolution of the radius $R$ for $k=0.35$, $l=0.5$, $C=0.292$ and $R(0)=2.923$.}
\end{center}
\label{fig:figure1}
\end{figure}
\begin{figure}[htbp!]
\begin{center}
\scalebox{0.6}{\includegraphics[angle=0]{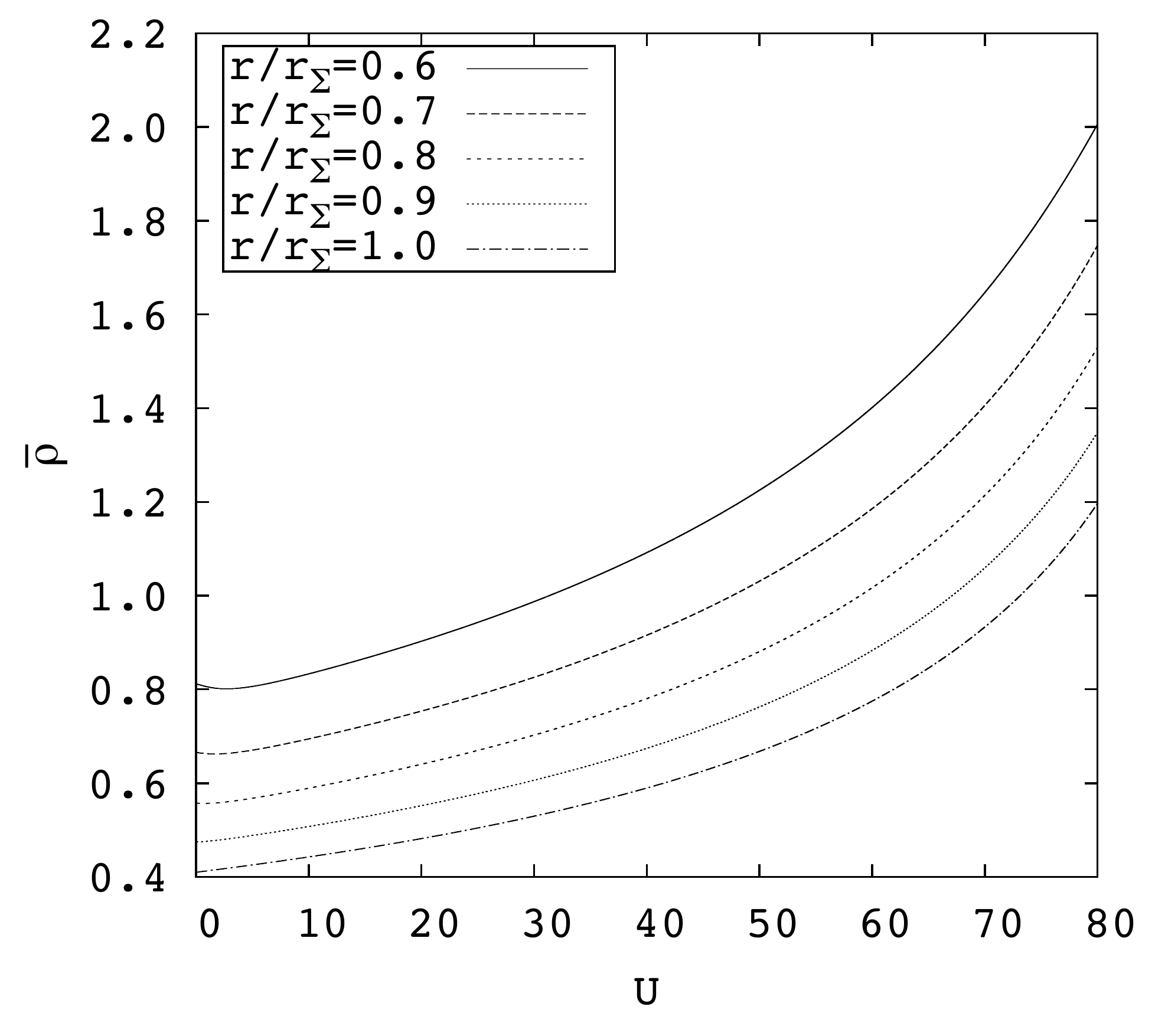}}
\caption{Evolution of the dimensionless energy density $\bar\rho =\tilde m_\Sigma(0)^2 \hat \rho$ (multiplied by $10^2$) for $k=0.35$, $l=0.5$, $C=0.292$ and $R(0)=2.923$.}
\end{center}
\label{fig:figure2}
\end{figure}
\begin{figure}[htbp!]
\begin{center}
\scalebox{0.6}{\includegraphics[angle=0]{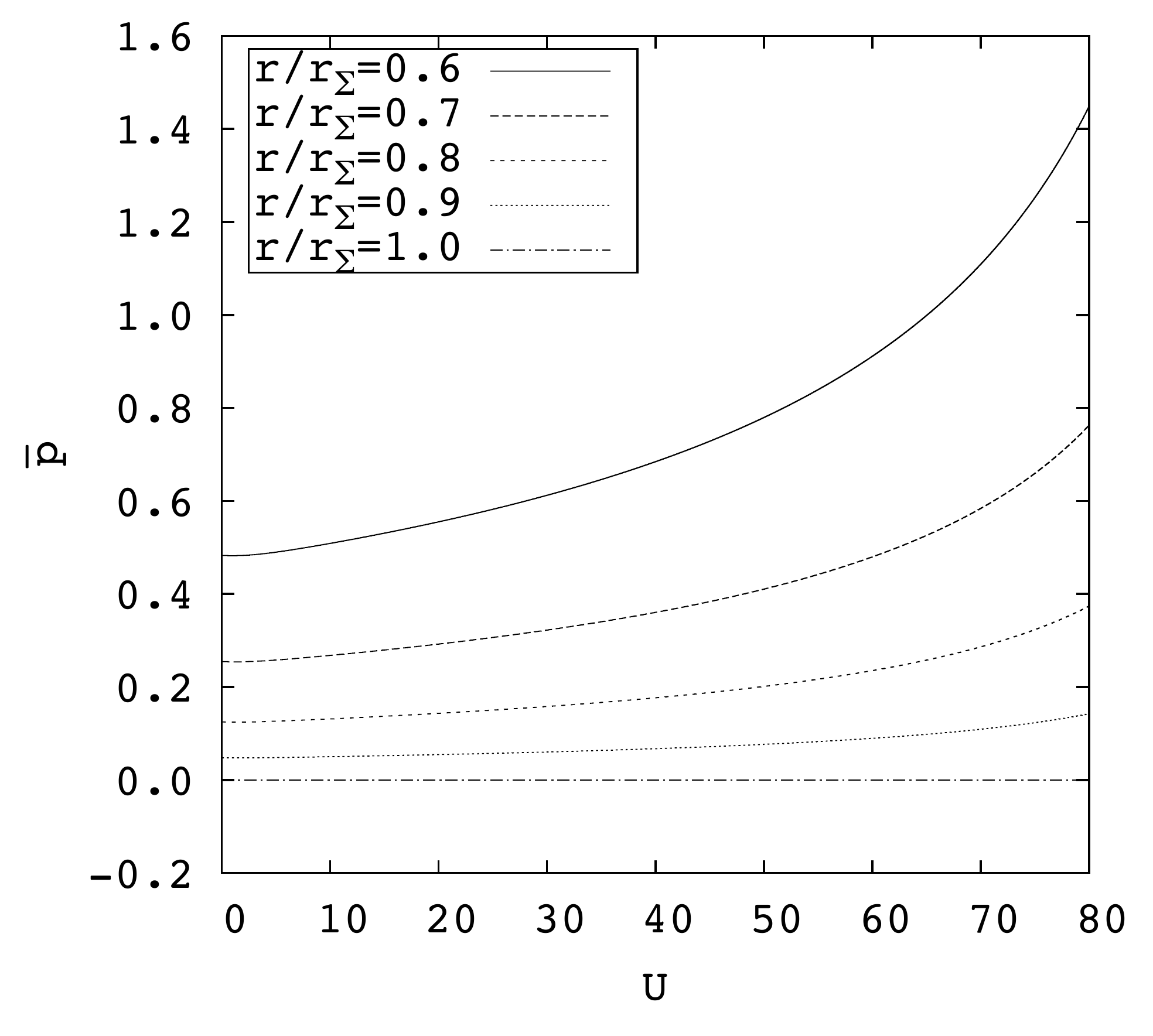}}
\caption{Evolution of the dimensionless radial pressure $\bar p=\tilde m_\Sigma(0)^2 \hat p$ (multiplied by $10^2$) for $k=0.35$, $l=0.5$, $C=0.292$ and $R(0)=2.923$.}
\end{center}
\label{fig:figure3}
\end{figure}
\begin{figure}[htbp!]
\begin{center}
\scalebox{0.6}{\includegraphics[angle=0]{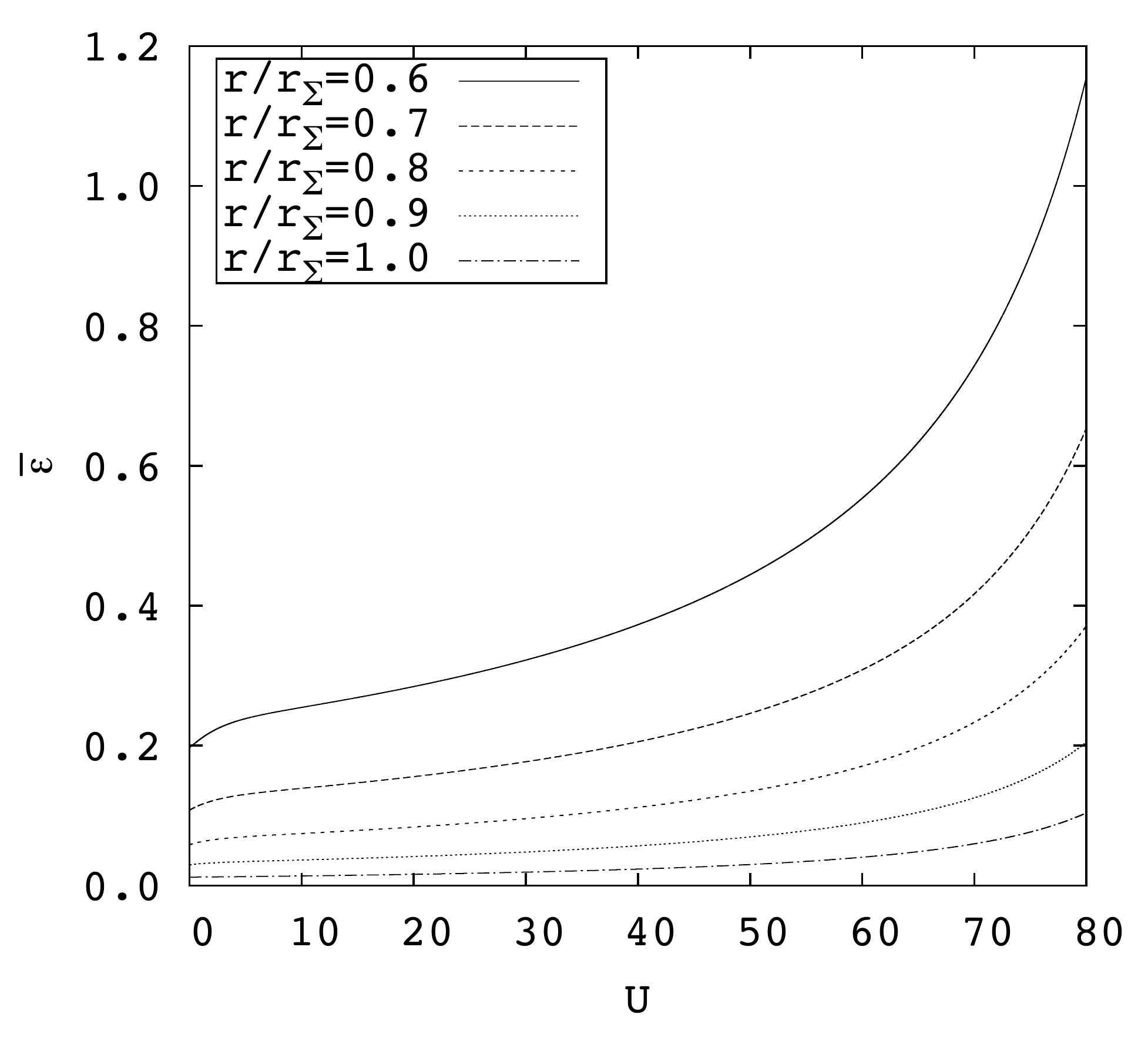}}
\caption{Evolution of the dimensionless energy flux $\bar\epsilon=\tilde m_\Sigma(0)^2 \hat \epsilon$ (multiplied by $10^2$) for $k=0.35$, $l=0.5$, $C=0.292$ and $R(0)=2.923$.}
\end{center}
\label{fig:figure4}
\end{figure}
\begin{figure}[htbp!]
\begin{center}
\scalebox{0.6}{\includegraphics[angle=0]{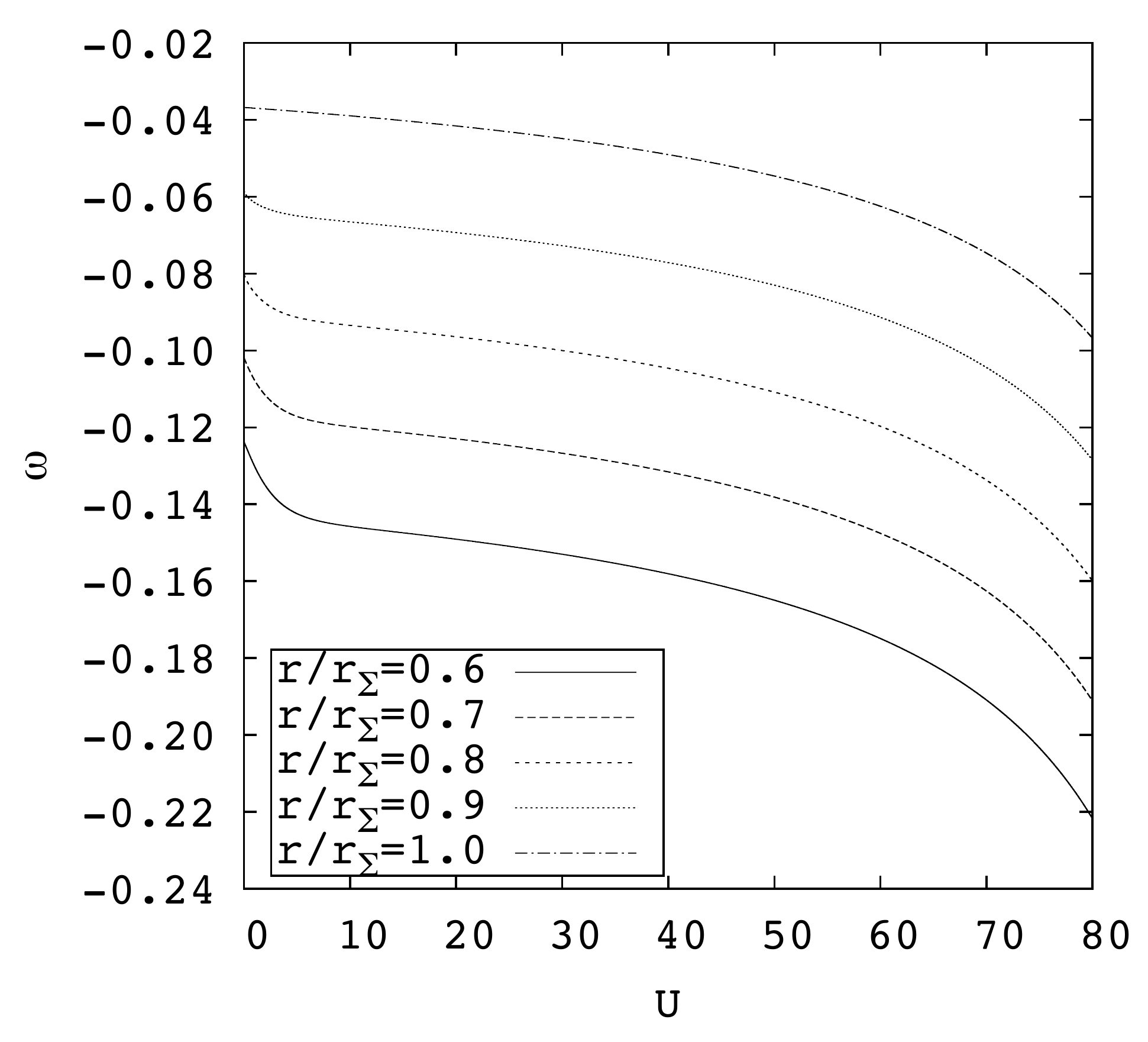}}
\caption{Evolution of the radial velocity $\omega$ for $k=0.35$, $l=0.5$, $C=0.292$ and $R(0)=2.923$.}
\end{center}
\label{fig:figure5}
\end{figure}
\begin{figure}[htbp!]
\begin{center}
\scalebox{0.6}{\includegraphics[angle=0]{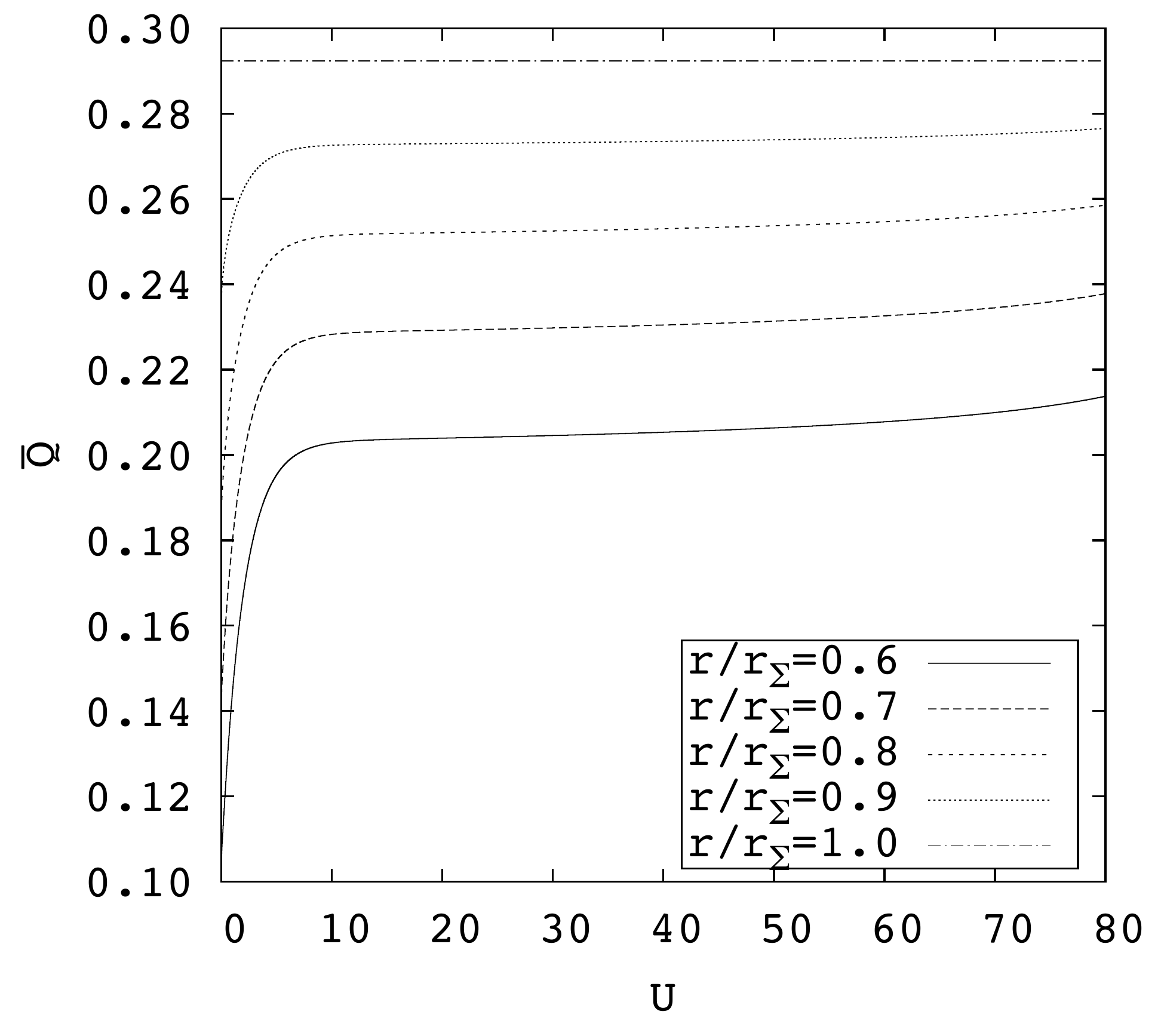}}
\caption{Evolution of the dimensionless charge function $\bar Q=\tilde m_\Sigma(0) Q$ for $k=0.35$, $l=0.5$, $C=0.292$ and $R(0)=2.923$.}
\end{center}
\label{fig:figure3}
\end{figure}
Clearly the metric functions and consequently the matter variables and electric charge are determined
up to the time dependent functions $\tilde m_\Sigma$ and $r_\Sigma$ for which we have the 
surface equations obtained from (\ref{one}) and (\ref{mv}) evaluated at $\Sigma$ 
\begin{equation}
\dot{\tilde m}_\Sigma=-4\pi r_\Sigma^2\epsilon_\Sigma\left(1-\frac{2\tilde m_\Sigma}{r_\Sigma}\right) + \dot r_\Sigma \frac{q^2}{2r^2_\Sigma}
\label{fse}
\end{equation}
and
\begin{equation}
\dot r_\Sigma=\left(1-\frac{2\tilde m_\Sigma}{r_\Sigma}\right)\frac{\omega_\Sigma}{1-\omega_\Sigma},
\label{sse}
\end{equation}
where $\epsilon_\Sigma=\hat\epsilon_\Sigma (1+\omega_\Sigma)/(1-\omega_\Sigma)$. 
Using the dimensionless variables
$$M=\frac{\tilde m_\Sigma\,\,\,\,\,}{\tilde m_\Sigma(0)}; \;\; R=\frac{r_\Sigma\,\,\,\,\,}{\tilde m_\Sigma(0)},$$
including the dimensionless time and electric total charge
$$U=\frac{u\;\;\;\;}{\tilde m_\Sigma(0)}; \;\; C=\frac{q\;\;\;\;}{\tilde m_\Sigma(0)},$$
the last two equations can be written as
\begin{equation}
\frac{dM}{dU}=-L\left(1-\frac{2M}{R}\right)+\frac{dR}{dU}\frac{C^2}{2R^2}
\end{equation}
and
\begin{equation}
\frac{dR}{dU}=\left(1-\frac{2M}{R}\right)\frac{\omega_\Sigma}{1-\omega_\Sigma},
\end{equation}
where $L=4\pi r_\Sigma^2\epsilon_\Sigma$.
From (\ref{sff}) we obtain
\begin{equation}
\omega_\Sigma=1-\frac{2(l+1)(1-2M/R)}{2(k+1)M/R-C^2/R^2}.
\label{omega}
\end{equation}
Now, from the field equation (\ref{four}) evaluated at $\Sigma$ we obtain $M$ as a function
of $R$,
\begin{equation}
M=\frac{1}{\psi}\left(\xi R-\frac{C^2}{R}\right),
\label{mass}
\end{equation}
where $\xi=(l+1)^2$ and $\psi=2\xi+k(4+3l+k)$
and consequently we get $L$ from (\ref{fse}) and (\ref{sse})
\begin{equation}
L=\left\{\frac{C^2}{R^2}\left[\frac{1}{2}+\frac{1}{\psi}\right]-\frac{\xi}{\psi}\right\}\frac{\omega_\Sigma}
{1-\omega_\Sigma}.
\end{equation}

Thus, we have one independent surface equation to integrate numerically if we specify initially in some way $M$ or $R$.  We have preference for (\ref{sse}), taking into account (\ref{omega}) and (\ref{mass}). 

We do not have, {\it a priori}, any restriction on values of $k$, $l$ and $C$. Only physical
criteria and expectations on the foregoing models determine our choices, as we illustrate in the following Section. For instance, any choice of values for  $k$, $l$ and $C$ have to be consistent with $-1<\omega < 1$, $\tilde m \ge 0$, $\hat \rho > 0$, $\hat \rho > \hat p$ for any light cone.

\section{Modeling}
As a first step for modeling we want to know the minimal radius
for a given set of $(k,l,C)$ imposing the conditions on the surface of the distribution.
It is easy to show that $\hat \rho_\Sigma \ge 0$ imposes the minimum radius: 
$$R^2\ge \frac{C^2}{\xi}\left\{\frac{\psi}{2(k+1)}+1\right\},$$
if $\omega_\Sigma>-1$. It is also important to note that any model based on self--similar solutions (\ref{eme})--(\ref{beta}) is singular at $r=0$. We specify an initial profile for the dimensionless electric charge function as
\begin{equation}
\bar Q(0)=C\left(\frac{r}{r_\Sigma}\right)^\mathcal{P},\label{Qi}
\end{equation}
where $\mathcal{P}$ is a free parameter.

We do a general survey for $k$, $l$ and $C$. There are values for which the distribution apparently is static. 
For instance, for $k=0.35$, $l\approx0.575$ and $C=10^{-3}$, the sphere has
$\omega_\Sigma\approx1.192\times 10^{-7}$ and $L\approx-4.172\times10^{-8}$, and stays there indefinitely. The situation remains the same for other values of total electric charge.

We opt for a general model with
$$k=0.35;\;\;l=0.5;\;\; C=0.292; \;\; R(0)=2.923.$$
In this case $M(0)\approx1$ and $\omega_\Sigma(0)\approx-0.037$. Thus, we integrate numerically the surface equation (\ref{sse}) using the fourth order Runge--Kutta method, constrained by (\ref{omega}) and (\ref{mass}). Choosing the free parameter $\mathcal{P}=2$ we integrate
numerically the equation (\ref{ce}) using finite differences. The procedure is straightforward and standard (see \cite{rbpr10} and references therein). It basically consists of using the Lax method (with the appropriate Courant--Friedrichs--Levy (CFL) condition). The conservation equation dynamics is restricted by the surface evolution. Once the charge function is advanced in time we can get any other physical variable. Thus the evolution of the whole distribution proceed up to the final time. For other choice of parameters and initial conditions the procedure is the same and the results displayed in figures 1--6 are representative. 
\begin{figure}[htbp!]
\begin{center}
\scalebox{0.6}{\includegraphics[angle=0]{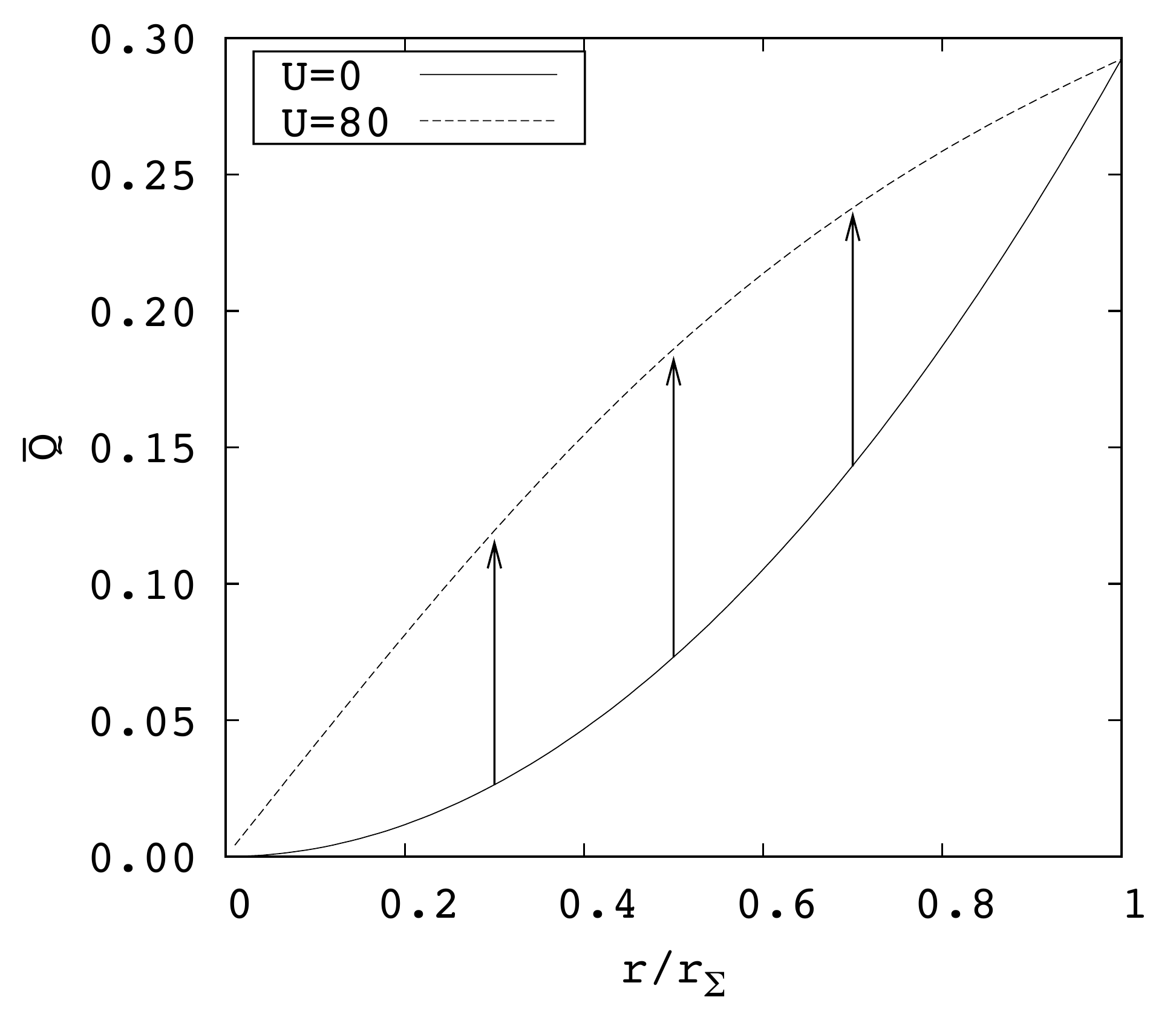}}
\caption{Dimensionless charge function $\bar Q=\tilde m_\Sigma(0) Q$ for $k=0.35$, $l=0.5$, $C=0.292$ and $R(0)=2.923$, as a function of comoving markers $r/r_\Sigma$ for initial and final time $U$. The electric charge redistributes in the interior space with time.}
\end{center}
\label{fig:figure3}
\end{figure}
\section{Discussion}
In this paper we consider electrically charged matter as anisotropic matter. We explore a dynamical model under the free streaming approximation and self--similarity within the source. The example we have shown is representative of many others varying the initial condition for the radius $R$, the self--similar parameters $k$, $l$, the total charge $C$. Only regions $0.6\le r/r_\Sigma \le 1$ satisfy physical conditions pointed out by end of Section 3. 

The main motivation for this work were previous results using the same system and solutions but a different transport mechanism, that is, the diffusion limit \cite{brrs07}. Heat flow makes the distribution evolve in a very different way: i) the electric charge halts the collapse; ii) the distribution becomes dust asymptotically in one special case; iii) the final state is oscillatory; iv) the electric charge is redistributed.

In the present case the electric charge does not change the fate of the gravitational collapse. 
The distribution evolves radiating a huge quantity of its mass ($\approx 45\%$) reaching relativistic velocities of collapse. In fact, inner regions are out of the physical domain.
Although the evolution looks catastrophic no evidence of black hole formation appears during the monitored evolution.

The only common feature for both transport mechanisms is the electric charge
redistribution, but in an opposite manner as we explain below. 

{We can read
   from figure 6 that for any time
   the interior electric charge, for any
   comoving space marker $r/r_\Sigma$, is always less
   than the total electric charge enclosed by the boundary surface.
   Therefore, the electric charge for inner regions can in fact increase (or decrease)
   by means a redistribution, conserving the total electric
   charge. In general, the electric charge grows from zero to the maximum value
   (the total) in all cases and, in the present case, as $(r/r_\Sigma)^{2}$
   initially and near $(r/r_\Sigma)^{4/5}$ later, conserving partially and
   totally the electric charge, as dictated by Eq. (7). It is also important to consider
      that redistribution occurs while the sphere is contracting.
   Figure 7 shows clearly these issues. In reference
   \cite{rbpr10} an analogous behavior is shown in other context,
   using other method of solution but with the same transport mechanism (see figures 15 and 16 there).}
   
{Why the differences found in the redistribution of
   charge depend on the transport mechanism? Clearly, the profiles of the flux energy density for the present case (figure 4) and
   the heat flux as reported in \cite{brrs07} (figure 4 there),
   are responsible for opposite radial velocity profiles (figure 5 in \cite{brrs07} and figure 5 here) and in consequence opposite electric
   charge redistribution (figure 6 in \cite{brrs07} and figure 6 here), conserving in all cases the electric charge.
   We mean by opposite that if the free streaming (heat flow) is lesser (greater) at the interior,
   the radial velocity is greater (negatively) at the interior, showing the opposite behavior for the
   diffusion mechanism, that is, lesser (negatively), all this for time--windowing where comparison applies. Thus, opposite electric charge redistribution means that between fixed extremes the inner electric charge per shell increases with evolution for the streaming out and diminishes for the diffusion transport mechanism.}

Generic numerical solutions for homothetic motion are currently under  consideration and will be reported elsewhere.

\section*{Acknowledgments} 
We are grateful to Carlos Peralta for his valuable reading and comments. {We thank the anonymous Referee for comments and questions raised during the review process to clarify results.}

\end{document}